\documentclass[twocolumn,amssymb, nobibnotes, aps, amsmath, amssymb, superscriptaddress,floatfix,]{revtex4-1}

\usepackage[utf8]{inputenc}
\usepackage[T1]{fontenc}
\usepackage{graphicx}
\usepackage{dcolumn}
\usepackage{bm}
\usepackage{newtxtext, newtxmath}  
\DeclareGraphicsExtensions{.pdf,.jpeg,.jpg, .png, .eps, .tiff}
\usepackage{epstopdf}
\usepackage{lipsum}
\usepackage{tabularx}
\usepackage{booktabs}
\usepackage{array}
\usepackage{natbib}
\usepackage[dvipsnames]{xcolor}
\usepackage{grffile}
\usepackage{boxhandler}
\usepackage{float}
\usepackage[normalem]{ulem}
\usepackage{xr-hyper} 
\usepackage{hyperref}
\usepackage{wasysym}
\usepackage{soul}
\usepackage{siunitx}
\usepackage[shortlabels]{enumitem}

\usepackage{xr}
\makeatletter

\newcommand*{\addFileDependency}[1]{
\typeout{(#1)}
\@addtofilelist{#1}
\IfFileExists{#1}{}{\typeout{No file #1.}}
}\makeatother

\newcommand*{\myexternaldocument}[1]{%
\externaldocument{#1}%
\addFileDependency{#1.tex}%
\addFileDependency{#1.aux}%
}

\myexternaldocument{si}
\myexternaldocument{response}

\renewcommand{\vec}[1]{\boldsymbol{#1}}
\newcommand{\av}[1]{\left\langle #1 \right\rangle}

\newcommand{\tens}[1]{\boldsymbol{\mathbf{#1}}}
\newcommand{\grad}{\vec{\nabla}}

\newcommand{\fig}[1]{\textbf{Fig.~\ref{#1}}}

\newcommand{\figSI}[1]{\textbf{Fig.~S{#1}}}
\newcommand{\eq}[1]{\textbf{Eq.~\ref{#1}}}
\newcommand{\EQ}[1]{\textbf{Equation~\ref{#1}}}

\newcommand{\movie}[1]{\textbf{movie~\ref*{#1}}}
\newcommand{\movies}[2]{\textbf{movies~{#1}-{#2}}}
\newcommand{\hardref}[1]{\textbf{#1}}

\newcounter{simovie}
\newcommand{\dummymov}[1]{\refstepcounter{simovie}\label{#1}}

\newcommand{\vel}{\vec{v}}
\newcommand{\vx}{v_x}
\newcommand{\vy}{v_y}
\newcommand{\act}{\zeta}
\newcommand{\actleng}{\ell_{\act}}
\newcommand{\visc}{\eta}

\newcommand{\Q}{Q}
\newcommand{\tensQ}{\tens{\Q}}

\newcommand{\lb}{\ell_B}
\newcommand{\perm}{\kappa}
\newcommand{\por}{\phi}

\newcommand{\darcyV}{U_{\rm{D}}}
\newcommand{\actdarcyV}{U_{\rm{AD}}}
\newcommand{\poiseuilleV}{U_{\rm{P}}}
\newcommand{\G}{G}
\newcommand{\J}{J}
\newcommand{\W}{W}

\newcommand{\flowpro}{\av{ \vx }_{x,t}}

\newcommand{\flowproInst}{\av{ \vx }_{x}}
\newcommand{\drift}{\av{\vx}_{x,y}}
\newcommand{\avgdrift}{\av{ \vx }_{x,y,t}}

\newcommand{\ke}{\av{ \vel^2 }_{x,y,t}}

\newcommand{\ue}{School of Physics and Astronomy, The University of Edinburgh, Peter Guthrie Tait Road, Edinburgh, EH9 3FD, United Kingdom}
\newcommand{\uc}{Niels Bohr Institute, University of Copenhagen, Blegdamsvej 17, Copenhagen, Denmark.}

\definecolor{pumpkin}{rgb}{1.0,0.4,0.0}
\definecolor{mygreen}{rgb}{0.0,0.55,0.3}
\definecolor{strawberry}{rgb}{1.0,0.0,0.5}
\definecolor{midnight}{rgb}{0.003921569,0.098039216,0.576470588}
\definecolor{saphire}{rgb}{0.0,0.196,0.372549}
\definecolor{crimson}{rgb}{0.75686,0,0.262745}
\definecolor{capri}{rgb}{0.0,0.768627,0.8745098}

\begin{document}

\title{Active Darcy's Law}
\author{Ryan R. Keogh}
\affiliation{\ue}
\author{Timofey Kozhukhov}
\affiliation{\ue}
\author{Kristian Thijssen}
\affiliation{\uc}
\author{Tyler N. Shendruk}
\email{t.shendruk@ed.ac.uk}
\affiliation{\ue}

\begin{abstract}
While bacterial swarms can exhibit active turbulence in vacant spaces, they naturally inhabit crowded environments. We numerically show that driving disorderly active fluids through porous media enhances Darcy's law. While purely active flows average to zero flux, hybrid active/driven flows display greater drift than pure-driven fluids. This enhancement is non-monotonic with activity, leading to an optimal activity to maximise flow rate. We incorporate the active contribution into an active Darcy's law, which may serve to help understand anomalous transport of swarming in porous media.
\end{abstract}

\maketitle

\begin{figure*}[tb]
    \centering
    \includegraphics[width=0.95\textwidth]{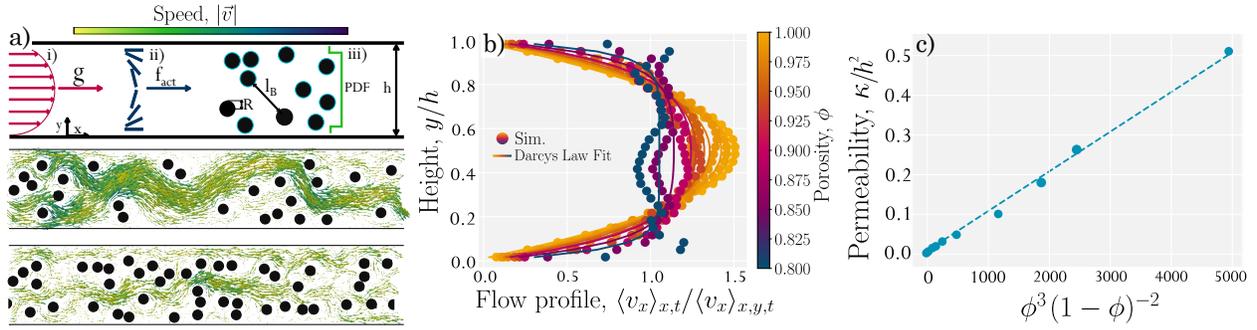}
    \caption{
        \textbf{Purely pressure-driven flow through porous media.} 
        \textbf{a)} 
        Top: Schematic of pressure-driven active nematic fluids within a obstacle-laden rectangular channel of height $h$. \textit{i)} Flow is driven by the pressure gradient $-G$. \textit{ii)} Extensile active nematics generate local active forces $\vec{f}_\text{act}$. \textit{iii)} Obstacles of radius $R$ are placed with a homogeneous probability distribution function (PDF), without overlaps with each other or the walls, creating a porous medium characterised by Brinkman pore size $\lb$. 
        Middle: Snapshot of passive flow ($\act=0$) due to a pressure gradient ($G=0.011$), coloured by speed at porosity $\por=0.87$. 
        Bottom: Snapshots of passive flow ($\act=0$) due to a pressure gradient ($G=0.011$) at porosity $\por=0.79$. 
        The colour bar shows the magnitude of flow speed ranging from $0.0\to0.3$ in MPCD units.
        \textbf{b)} Normalised flow profiles $\flowpro$ across the channel, fit by \hardref{Eq. S(8)}
         (solid lines). 
        \textbf{c)} Permeability $\perm=\lb^2$ from fits in (b) grow linearly with the Kozeny–Carman factor $\por^{3}(1-\por)^{-2}$ (dashed line).
    }
    \label{fig:passFlow}
\end{figure*}

\begin{figure}[h!]
    \centering
    \includegraphics[width=0.8\columnwidth]{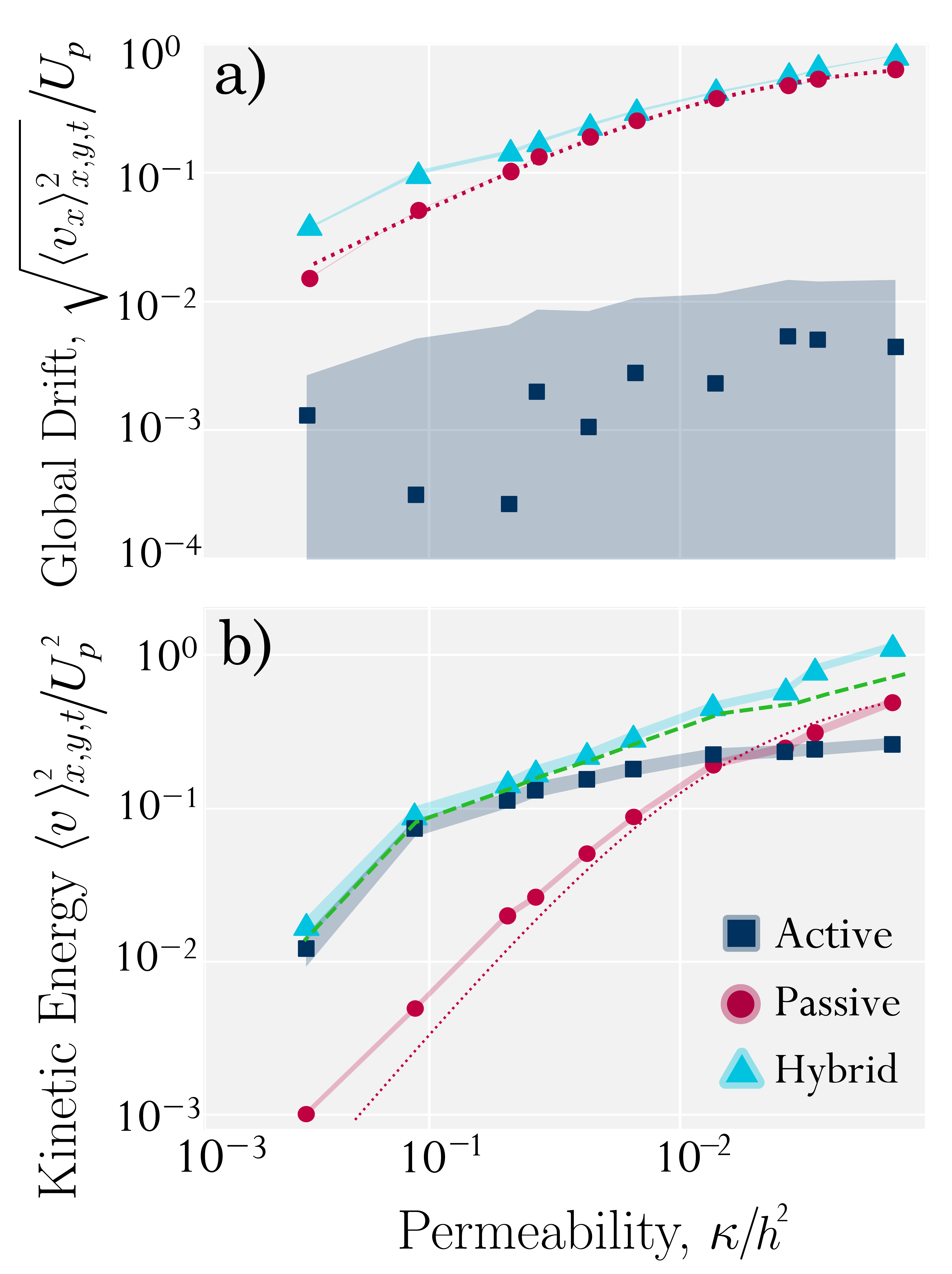}
    \caption{
        \textbf{Global averaged flows.} 
         {\color{crimson}$\boldsymbol{\circ}$}---Purely pressure-driven ($\act=0$; $G=0.011$), 
         {\color{saphire}$\boldsymbol{\square}$}---Purely active ($\act=0.08$ ($\actleng\simeq10$); $G=0$), and 
         {\color{capri}$\boldsymbol{\triangle}$} Hybrid active/pressure-driven ($\act=0.08$; $G=0.011$). 
        Shaded regions denote standard error. 
        \textbf{a)} Time averaged global drift $\avgdrift$ non-dimensionalised by Poiseuille flow $\poiseuilleV$ for $\G=0.011$.
        The pressure-driven case agrees with Darcy's law (\hardref{Eq. S(9)}
        ; dotted line), while the purely active case exhibits zero drift. 
        \textbf{b)} Kinetic energy density $\ke$. 
        Pressure-driven flow is consistent with theory (dotted red line) and hybrid flow coincides with the sum of the pure cases (dashed green line). 
    }
    \label{fig:globalAvg}
\end{figure}

Active fluids spontaneously flow because energy is locally injected by inherently out-of-equilibrium particles~\cite{shankar2022}, such as motile cells~\cite{beer2019statistical} and exhibit collective dynamics on scales many times larger than individual cells~\cite{Alert2022}. 
Such coherent flows---uncorrelated at large spatio-temporal scales---describe many cellular systems, including swarming bacteria~\cite{nijjer2021,Liu2021density,Aranson2022} and disorderly bacterial turbulence~\cite{wensink2012meso,Ariel2018,Peng2021}. 
These flows have been studied in engineered geometries, e.g. channels~\cite{thampi2022,Henshaw2023}, cavities~\cite{wioland2013confinement,Lushi2014,liu2021}, annuli~\cite{Wioland2016directed,creppy2016}, connected voids~\cite{wioland2016ferromagnetic}, pillar arrays~\cite{nishiguchi2018,reinken2020}, ratchets~\cite{di2010bacterial}, slits~\cite{mackay2020darcy} and periodic obstacles~\cite{pietzonka2019autonomous, reinken2020organizing, zhang2020oscillatory, reichhardt2020directional, chen2022microbial}.
However, many naturally occurring examples of collective bacterial motion arise in heterogeneous environments, such as porous media~\cite{ranjbaran2020mechanistic,wu2020medical, dentz2022dispersion}. 

Individual self-propelled particles in the vicinity of obstacles have been studied, including bacteria~\cite{bhattacharjee2019confinement, krishnamurthi2022interactions, lee2021influence, perez2021impact}, active Brownian particles~\cite{jakuszeit2019diffusion, moore2023active, modica2023boundary} and active polymers~\cite{mokhtari2019dynamics, kurzthaler2021geometric, theeyancheri2022migration}. 
Heterogeneity can have profound effects, on individual and collective dynamics, including clogging~\cite{coyte2017microbial, bhattacharjee2019confinement, shrestha2023bacterial, dehkharghani2023self}, rectified conductivity~\cite{waisbord2021fluidic}, boundary migration~\cite{rismani2018migration}, intermittency~\cite{Secchi2016}, redistribution~\cite{scheidweiler2020trait,teo2022microfluidic}, chemotactic response~\cite{bhattacharjee2021chemotactic,bhattacharjee2022chemotactic,gao2023chemotaxis}, topological flocking~\cite{rahmani2021topological} and other novel collective movements~\cite{bhattacharjee2019bacterial,engelhardt2022novel,amchin2022influence,BenDor2022}. 
Indeed, the generality of anomalous motion in heterogeneous media~\cite{alim2017local,anbari2018microfluidic,wu2020nanoparticle,Shende2022,fan2022anomalous} suggests porosity crucially alters active collective transport~\cite{kumar2022transport}.

Here, we study the cooperative effect of disorderly active flows on pressure-driven transport through porous materials of fixed obstacles (\fig{fig:passFlow}a). 
We find that, when biased due to weak external pressure gradients, activity positively enhances global drift, even in the active turbulence limit. 
This effect endures as the number density of obstacles is increased, resulting in undiminished kinetic energy despite increased dissipative drag---suggesting active flows autonomously fill the available porous length scales. 
The active auxiliary flux increases with pressure gradient and is non-monotonic with activity, possessing an optimal activity.  
We conclude that the flux of active fluids through porous media are described by a modified Darcy's law. 


\begin{figure*}[tb]
    \centering
    \includegraphics[width=0.95\textwidth]{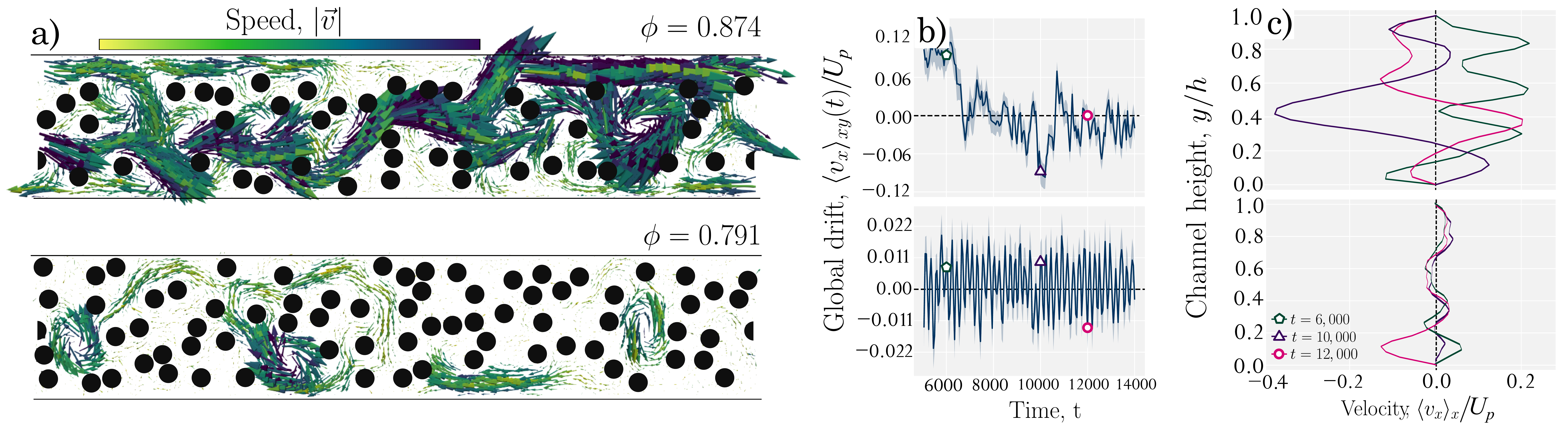}
    \caption{
        \textbf{Purely active nematic ($\act=0.08$) through porous media.} 
        Top row: Porosity $\por=0.87$ ($\perm/h^2=6.4\times10^{-3}$). Bottom row: $\por=0.79$ ($\perm/h^2=2.3\times10^{-3}$). 
        \textbf{a)} Snapshots of pressure-gradient-free ($G=0$) active nematic flow. The colour bar represents the magnitude of flow speed in MPCD units from $0.0\to0.3$.
        \textbf{b)} Instantaneous global drift $\drift(t)$ non-dimensionalised by Poiseuille flow $\poiseuilleV$ for $\G=0.011$ for systems shown in (a). 
        For the more dilute porous medium (top), the drift direction persists over time, whereas the denser system (bottom) exhibits more rapid fluctuations. 
        Highlighted times: {\color{OliveGreen}$\pentagon$}---$t=6\times10^3$, {\color{Violet}$\triangle$}---$10\times10^2$, and {\color{RubineRed}$\circ$}---$12\times10^3$. 
        \textbf{c)} Instantaneous flow profiles $\flowproInst(y;t)$ highlighted times in b). 
      }
    \label{fig:actFlow}
\end{figure*}

While individual swimming bacteria move with polar self-propulsion, their dipolar hydrodynamic and head-tail-symmetric steric interactions are nematic, making active nematics an appropriate minimal model~\cite{Baskaran2008,saintillan2015,peshkov2014}, in agreement with experiments~\cite{Nishiguchi2017,li2019,Peng2021,Liu2021density}. 
To model active nematic fluids within a complex geometry, we employ active-nematic Multi-Particle Collision Dynamics (AN-MPCD~\cite{si}), a recent mesoscopic method~\cite{kozhukhov2022}. 
AN-MPCD simulates linearised fluctuating nematodynamics with isotropic viscosity and elasticity~\cite{Hijar2019}, and activity $\act$ modelled via local multi-particle force dipoles~\cite{kozhukhov2023-ANSig}
The competition between elasticity and activity results in an active length scale $\actleng$ (\figSI{1}). 
The corresponding continuum equations correspond to low Reynolds number flows with linearised nematic elasticity and an active stress term \cite{doostmohammadi2018active,shendruk2017dancing}. 
Systems of motile bacterial exhibit density fluctuations~\cite{Zhang2010,liu2021}, as does the particle-based AN-MPCD approach (\figSI{2}), in contrast to continuum models that assume incompressibility. 
We focus on flow-aligning nematics with extensile activity $\act$ and a weak external forcing $-\G=\left(\grad P\right)\cdot\hat{x}$ due to a pressure gradient $\grad P$ down the channel $\hat{x}$ (\fig{fig:passFlow}a), which breaks the directional symmetry of bacterial turbulence~\cite{wensink2012meso}. 
Values are reported in MPCD units~\cite{si} and unless otherwise stated, $\G=0.011$ and $\act=0.08$ $\actleng \simeq 10$). 
All other parameters are chosen to match previous studies characterising the numerical approach~\cite{kozhukhov2022}. 
AN-MPCD is ideal for simulating active flows around randomly-placed off-lattice obstacles with a broad distribution of voids. 
The active fluid and porous medium are confined within a 2D channel with height$h=30$ and length $L=150$ with impermeable, no-slip walls with free-nematic anchoring (\fig{fig:passFlow}a), which models well-defined experimental set-ups~\cite{anbari2018microfluidic}. 
Finite-size effects (\figSI{3}) and alternative anchoring conditions (\figSI{4}) are considered in the supplementary materials. 
The porous medium is formed of impermeable, immobile, circular obstacles of radii $R=2$ with the same boundary conditions as the walls, producing isotropic porosity $\por\in[0.67,1.0]$, where $\por=1.0$ is an empty channel and $0.67$ is likely to be impermeable~\cite{si}. The random obstacles are homogeneously distributed (\fig{fig:passFlow}a) but overlaps are not permitted. 
Components of the velocity field $\vel = \vx\hat{x} + \vy\hat{y}$ are averaged temporally $t$, longitudinally $x$ and/or transversely $y$, denoted as $\av{\cdot}_{x,y,t}$ to measure flow profiles and global fluxes. 


Before considering cooperative effects, we quantify the properties of the porous medium.  
In the limit of zero activity and large porosity ($\por\to1$), the flow is relatively unobstructed and the mean velocity profile $\flowpro(y)$ is parabolic (\fig{fig:passFlow}b; yellow). 
As the porosity is decreased, the greater number of obstacles slows the flow (\fig{fig:passFlow}a; middle $\phi=0.87$) and broadens the profiles (\fig{fig:passFlow}b; pink). 
At low porosity, the effect of the walls is minimal compared to obstacle drag (\fig{fig:passFlow}a; bottom $\phi=0.79$), producing plug-like flow (\fig{fig:passFlow}b; blue). 
The flow profiles follow the Brinkman equation~\cite{durlofsky1987analysis} $\left[ \partial_{yy}-\perm^{-1}\right]\av{\vx}_{x,t}({y}) = -\G/\visc$ for viscosity $\visc$ and interstitial permeability $\perm$~\cite{si}, indicating that anisotropic properties can be absorbed into the Brinkman terms (\fig{fig:passFlow}b). 
At the lowest porosity, systematic deviations, in the form of limited non-monotonic shoulders in the near-wall region appear (\fig{fig:passFlow}b; blue), due to steric inaccessibility of obstacles. 
From the fits, $\perm$ and corresponding pore size $\lb=\sqrt{\perm}$ (or Brinkman length) are found (\fig{fig:passFlow}c). 
The resulting permeability obeys the Kozeny–Carman relationship $\perm = c \por^3(1-\por)^{-2}$ with $c=0.091\pm0.004$. 

As permeability $\perm$ decreases so does global drift $\avgdrift$ (\fig{fig:globalAvg}a). 
In obstacle-laden channels ($\lb \ll h$), the drift is linearly dependent on $\perm$, which follows from Darcy velocity $\darcyV \equiv \G\perm/\visc$~\cite{si}; while in obstacle-free channels ($h\ll\lb$), the drift saturates to the expectation for Poiseuille flow $\poiseuilleV \equiv \G h^2/12\visc$. 
Both are encompased by $\avgdrift = \darcyV\W\left(h/\lb\right)$, where the correction due to channel confinement $\W\left(h/\lb\right)=1-\left( 2\lb/h \right)\tanh{\left( h/2\lb \right)}$ depends only on the dimensionless ratio $h/\lb$ (\fig{fig:globalAvg}a; dotted line~\cite{si}). 
Similarly, the kinetic energy density $\ke = \darcyV^2 \left[ 2+\cosh{\left(h/\lb\right)}-\left(3\lb/h\right)\sinh{\left(h/\lb\right)} \right] / \left[1+\cosh{\left(h/\lb\right)}\right]$ \cite{si} is in good agreement with the simulations  (\fig{fig:globalAvg}b). 


In contrast, purely active turbulent flows do not drift down the channel~\cite{wensink2012meso}. 
Even at lower activities, where spontaneous symmetry breaking results in unidirectional active flow, global drift averages to zero over multiple simulations~\cite{hardouin2019reconfigurable}.
We focus on activities which exhibit isotropic active turbulence in an obstacle-free channel ($h\gg\lb$)~\cite{doostmohammadi2017onset} (see \movie{dmov:Empty-turbulence}). 
As $\por\to1$, the channel height $h$ competes with the active length scale $\actleng$~\cite{hemingway2016correlation} to determine the spatiotemporal structure of the active flow~\cite{shendruk2017dancing,hardouin2019reconfigurable,keogh2022helical,head2023}, which result in vortex lattices ~\cite{shendruk2017dancing} recently observed as the first mode in suspensions of bacteria~\cite{Henshaw2023}. 
However, as obstacles are added, the confinement length switches from $h$ to pore size $\lb$. 
This causes the dynamics to transition from active turbulence to local unidirectional flows (\fig{fig:actFlow}a; top and \movie{dmov:HighPhi-streaming}). 
There are well defined paths in which unidirectional flow might arise, though spontaneous symmetry breaking might produce a stream of unidirectional flow moving in the opposite direction along other paths, or a recirculating vortex might become trapped in a void (\fig{fig:actFlow}a; top). 

The global drift $\drift(t)$ quantifies the instantaneous new flux, and can persist for finite durations (\fig{fig:actFlow}b; top).
Three instances illustrate different flow structures (\fig{fig:actFlow}c; top). 
First (\fig{fig:actFlow}c; top green), the instantaneous flow profile $\flowproInst(y,t)$ exhibits a significant net flux in the $+\hat{x}$ direction, despite opposing flow in the vicinity of the $y=0$ wall. 
This net flux $\drift(t)$ persists (\fig{fig:actFlow}b; top). 
The second moment (\fig{fig:actFlow}c; top purple) exhibits strong instantaneous flux $\drift(t)$ in $-\hat{x}$ direction; however, it is short-lived (\fig{fig:actFlow}b; top). 
The third time (\fig{fig:actFlow}c; pink) illustrates that $\drift(t)$ can be near-zero, despite non-zero flow in different regions of the channel. 

While large porosity can allow otherwise turbulent active flows to instantaneously possess net global drift (\fig{fig:actFlow}a; top), lower porosity hinders instantaneous drift (\fig{fig:actFlow}a; bottom). 
In lower porosity, pore-entrapped vortices are more frequent and long-lived (\fig{fig:actFlow}a; bottom and \movie{dmov:LowPhi-trapped}). 
Furthermore, system-spanning streams are less likely and net fluxes become fleeting and noisier (\fig{fig:actFlow}b; bottom), while the instantaneous flow profiles $\flowproInst(y,t)$ approach zero across the channel (\fig{fig:actFlow}c; bottom).
Although there are localised flows, drag on the obstacles generally dominates. 

Ensemble averaged active drift is $\avgdrift=0$, due to the spontaneous symmetry breaking in an absence of pressure gradients (\fig{fig:globalAvg}a).
However, despite zero net global flux in the purely active systems, there is significant global kinetic energy $\ke$ (\fig{fig:globalAvg}b). 
Whereas kinetic energy drops rapidly with decreasing permeability $\perm$ for pressure-driven flows (\fig{fig:globalAvg}b; red), active systems maintain higher kinetic energy at low permeability (\fig{fig:globalAvg}b; blue). 
For $\act=0.08$, the kinetic energy of the active flow is greater than the pressure-driven flow for $G=0.011$ when $\perm/h^2\lesssim0.11$.
This is because active forces still generate locally coherent flow between obstacles and pores with sizes comparable to the active length scale $\actleng$ trap vortices, allowing localised self-sustaining recirculations. 


While purely active systems have zero flux, this is not true of hybridised flows that are both active and externally biased (\figSI{5}). 
As in the purely activity case, activity $\act$ is chosen such that active flows are in the turbulent regime without obstacles. 
The pressure gradient $\G$ is sufficiently high that activity acts as a perturbation to the driven case, but not so high that flow alignment is enforced. 
The kinetic energy density approximately corresponds to the sum of the two pure cases (\fig{fig:globalAvg}b). 
At small pore sizes, the active kinetic energy of the hybrid system is close to that of the purely active flow. 
The hybrid flow begins to differ from the purely active case around $\perm^*/h^2\simeq0.022$. 
Here, the ratio of the Brinkman length and active length scale is near unity ($\lb^*/\actleng \sim 4.5/10$). 

\begin{figure*}[tb]
    \centering
    \includegraphics[width=0.95\textwidth]{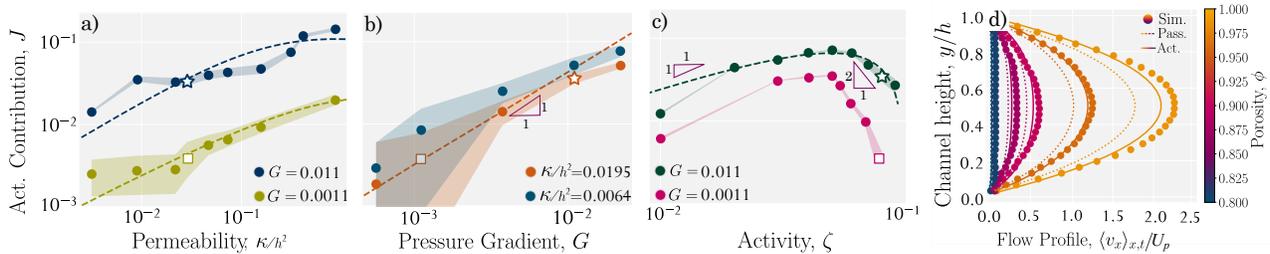}
    \caption{
        \textbf{Active contribution $\J$ to hybridised flow.} 
        $\bigstar$ and $\blacksquare$ symbols denote the same data sets across panels, dashed lines represent fits, and shaded regions are the standard error.
        \textbf{a)} Dependence of $\J$ on permeability $\perm$ with $\act=0.08$. 
        Dashed lines predicted from \eq{eq:actContr} with $\act^*$ fit of \eq{eq:actContr} to the data in (c). 
        \textbf{b)} Linear dependence of $\J$ on pressure gradient $\G$ for $\perm/h^2=1.9\times10^{-2}$ and $6.4\times10^{-3}$ with $\act=0.08$. 
        Dashed lines predicted from \eq{eq:actContr} with $\act^*$ fit of \eq{eq:actContr} to the data in (c). 
        \textbf{c)} Non-monotonic dependence of $\J$ on $\act$ with a maximum at $\act^*=0.05$ for $\perm/h^2=1.9\times10^{-2}$ ($\por=0.93$) with $\G=0.011$ and $0.0011$. 
        For $\G=0.011$, at low activity, $\J\sim\act$, while $\J\sim\act^{-2}$ at high activity. 
        Fit of \eq{eq:actContr} with $\act^*=0.05\pm0.02$ shown as dashed yellow line. 
        \textbf{d)} Hybrid flow profiles normalised by empty channel flow $U_p$ compared with passive and active Darcy's laws. 
        Simulated flow profiles (dots) for $\G=0.011$ compared to passive (dotted lines; \hardref{Eq. S(8)}
        ) and active (solid lines; \eq{eq:activeDarcy}) Darcy's law predictions. 
    }
    \label{fig:diffJ}
\end{figure*}

The increased kinetic energy acts as an auxiliary driving force, assisting the pressure gradient to conduct fluid through the porous space. 
The global flux shows an augmented drift when compared to either the purely driven or purely active cases (\fig{fig:globalAvg}a). 
Though the purely active flow is disorderly, in the hybrid case local active energy injection enhances the flux. 
To study the active contribution to the drift, we measure the difference $\J = \avgdrift^H-\avgdrift^P$, where $\avgdrift^H$ is the global drift of the hybrid case and $\avgdrift^P$ is the purely pressure-driven case. 
We find $\J>0$ in all instances (\fig{fig:diffJ}), which reveals that even disorderly active flows enhance pressure-driven flux in porous systems. 
The active enhancement $\J$ increases linearly with $\perm$ at low permeability then saturates, going as $\J\left(\kappa\right) \sim \perm \W(h/\sqrt{\kappa})$ (\fig{fig:diffJ}a). 
This is the same dependence as Darcy's law when going from a crowded channel to an obstacle-free channel (\fig{fig:globalAvg}; red circles). 
Similarly at sufficiently small activity, the active contribution increases linearly with pressure gradient $\J\sim\G$ (\fig{fig:diffJ}b), suggesting the enhancement requires a coupling to the external biasing --- the stronger the pressure gradient, the more the activity can boost the flux. 
Pressure gradients generate directed deformations to the orientation field $\tensQ$, which induce an auxiliary active forcing $f^{\rm{act}} \sim \langle{\partial_y^2}\Q\rangle \sim G\zeta$~\cite{walton2020pressure}. 
This approximation only holds in the limit $G < \act/\lb$, since flow-alignment eventually suppresses divergence of $\Q$ and thus the active auxiliary forcing slows at larger pressure gradients. 
this approximation suggests the active contribution is linear with activity. 

Indeed, varying activity $\act$ demonstrates that the enhancement $\J(\act)\sim\act$ increases linearly at sufficiently low activities (\fig{fig:diffJ}c; \movie{dmov:MidPhi-lowAct}). 
However, the auxiliary flux reaches a maximum value at $\act^*=0.05$ (\movie{dmov:MidPhi-peakAct}), then decreases (\fig{fig:diffJ}c), indicating there is an optimal activity for enhancing porous transport that is independent of the biasing pressure gradient. 
The maximum active contribution occurs when the ratio of the Brinkman and active length scales approach unity ($\lb/\actleng^* \sim 4.5/15$). 
For small activities (below the optimal value $\act < \act^*$), $\actleng$ is larger than the characteristic pore size---disorderly active turbulence does not occur within the pores.
Active flows are unidirectional and laminar-like within the pores; hence, $\J\sim\act$. 
However for $\act>\act^*$, active turbulence is possible within the pores and collective flows become uncorrelated even on scales smaller than $\lb$. 
The non-laminar uncorrelated turbulence generates additional kinetic dissipation and active vortices that are not conducive to flux (\movie{dmov:MidPhi-highAct}), causing the enhancement to decline as $\J\sim\act^{-2}$ (\fig{fig:diffJ}c). 
In active nematics, much of the kinetic energy is contained with the vortices and so the additional dissipation scales with the change of enstrophy $f^{\rm{disp}} \sim -\zeta^2$~\cite{giomi2015geometry}.

Having obtained the active auxiliary contribution as a function of its dependencies, the full effect of local activity on the global flux is
\begin{align}
    \J(\perm,\act,\G) &= \darcyV \ \left[\frac{\act}{2\act^*}-\left(\frac{\act}{2\act^*}\right)^2 \right]  \W\left(\frac{h}{\lb}\right),
    \label{eq:actContr}
\end{align}
where $2\act^*$ is the optimal activity at which the active length scale $\actleng$ matches the Brinkman length $\lb$. 
The active contribution only depends directly on the Darcy velocity and dimensionless ratios.
\EQ{eq:actContr} is linear in pressure gradient for $\G\lb/\act < 1$, nonlinear-but-monotonic with permeability and non-monotonic in activity, with a maximum where $\actleng=\lb$. 
The finite size of the channel enters through the factor of $\W$~\cite{si}. 
The active contribution in \eq{eq:actContr} is fit as a function of activity for a single pressure gradient (\fig{fig:diffJ}c; $\G=0.011$), with $\act^*$ the only fitting parameter since $\darcyV$ and $\W$ are given by the permeability from \fig{fig:passFlow}. Having fit $\act^*$ for a single system, the predicted enhancement as a function of $\G$ and $\perm$ is seen to agree with all other simulations without any further fitting (\fig{fig:diffJ}a-b). 
The fit is accurate for activities above the pressure-dominated limit ($\G\lb/\act > 1$), where and flow-alignment suppresses active forcing. 
Hence, activity enhances transport in porous channels via an effective active Darcy velocity 
\begin{align}
    \actdarcyV &= \frac{\G\perm}{\visc} \left[ 1+\frac{\act}{2\act^*}-\left(\frac{\act}{2\act^*}\right)^2 \right] ,
    \label{eq:activeDarcy}
\end{align}
which recasts the Brinkman equation and channel-averaged drift into active Brinkman and drift equations by substitution of $\darcyV\to\actdarcyV$. 
This prediction captures the enhancement of the flow profiles across the channel without additional fitting (\fig{fig:diffJ}d). 


Here, we studied the auxiliary contribution of collective bacterial motion on fluid transport through porous media. 
Our results show that activity enhances transport of pressure-driven fluids, even in the limit of bacterial turbulence. 
The pressure gradient breaks the symmetry of disorderly active flows, which enhances the total transport properties. 
Optimal activity for maximising the flux arises from competition between the characteristic active and porous length scales. 
By measuring the active contribution as a function of permeability, pressure gradient and activity, we discover an active version of Darcy's law. 
While the presented results are of active fluids and porous media confined within 2D channels, simulations of larger systems with periodic boundary conditions on all sides (\movies{7}{8}) confirm the proposed active Darcy's law (\figSI{6}). 
Through $\act^*$, the active Darcy equation can be written as $\actdarcyV/\darcyV = 1 + \left(\lb/\actleng\right)^2/2 - \left(\lb/\actleng\right)^4$/4. 

This expands on recent work studying driven active fluids in empty channels~\cite{mackay2020darcy,walton2020pressure} and previous description of activity lowering the apparent viscosity~\cite{cates2008shearing,giomi2010sheared,saintillan2018rheology,loisy2018active,martinez2020combined} to complex disordered environments. 
While the active Darcy's law presented here could be interpreted as reducing the apparent viscosity, raising the effective permeability or augmenting the pressure gradient, none of these interpretations make clear the physical mechanism of weak pressure gradients biasing the local force densities $\sim \act/\lb$ within pores to point in the same direction, nor do they account for an optimal value for active auxiliary forcing when the active length scale is comparable to pore size. 

The proposed active Darcy's law is particularly relevant for soil-associated bacteria as they collectively colonise plant roots within the rhizosphere~\cite{Venieraki2016}.
Indeed, model organisms for active turbulence, including \textit{Serratia marcescens} and \textit{Bacillus subtilis}, are common rhizobacteria, with characteristic bacteria turbulence scales $\actleng \sim 30\si{\micro\meter}$~\cite{Aranson2022,Sokolov2012,Sokolov2007,nishiguchi2018,li2019,ran2021}. 
These $\actleng$ are comparable to characteristic pore sizes within rhizospheric soils with permeability $\perm \sim 150\si{\micro\meter^2}$~\cite{Koebernick2017} and $\lb \sim 20\si{\micro\meter}$~\cite{Koebernick2019}. 
Our results indicate that $\lb/\actleng \sim 1$ is precisely the condition to maximise the active enhancement to the drift. 
Active transport enhancement could be utilised in future research as a framework to understand anomalous transport of nutrients by soil bacteria, as well as provide an approach for controlling active flows akin to continuous friction~\cite{Koch2021,Thijssen2021,MartinezPrat2021}. 
Previous work on dilute bacteria dynamics has highlighted the role of transient trapping~\cite{bhattacharjee2019bacterial,bhattacharjee2019confinement} and escape from cavities ~\cite{wu2021}, and dead-end pores~\cite{bordoloi2022structure}, and future work considering dense collective dynamics within random networks of cavities and dead-ends rather than obstacles may be fruitful. 
While we simulated immobile obstacles, future studies could consider how collective flows in turn modify deformable surroundings. 
Active clogging, erosion and infiltration may reveal much about the role of motile microbes as microecosystem engineers~\cite{Brussaard1997}. 

\section*{Acknowledgements}
This research has received funding (T.N.S. and K.T.) from the European Research Council under the European Union’s Horizon 2020 research and innovation programme (Grant Agreement Nos. 851196 and 101029079). 
For the purpose of open access, the author has applied a Creative Commons Attribution (CC BY) licence to any Author Accepted Manuscript version arising from this submission. 

\bibliography{references}

\dummymov{dmov:Empty-turbulence}
\dummymov{dmov:HighPhi-streaming}
\dummymov{dmov:LowPhi-trapped}
\dummymov{dmov:MidPhi-lowAct}
\dummymov{dmov:MidPhi-peakAct}
\dummymov{dmov:MidPhi-highAct}

\end{document}


\title{Supplementary material\\Active Darcy's Law}
\author{Ryan R. Keogh}
\affiliation{\ue}
\author{Timofey Kozhukhov}
\affiliation{\ue}
\author{Kristian Thijssen}
\affiliation{\uc}
\author{Tyler N. Shendruk}
\email{t.shendruk@ed.ac.uk}
\affiliation{\ue}

\maketitle




\section{Active-Nematic Multi-Particle Collision Dynamics}
\label{sctn:mpcd}
To simulate active nematic fluids within complex geometries, we use Active-Nematic Multi-Particle Collision Dynamics (AN-MPCD)~\cite{kozhukhov2022}, which allows off-lattice obstacles to be placed randomly throughout space with a broad distribution of voids in the porous medium. 
The active suspension is discretised into $N$ point particles, and each particle $i$ possesses a mass $m$ with instantaneous position $\mpcdpos(t)$, velocity $\mpcdvel(t)$ and orientation $\vec{n}_i(t)$. 
By simulating an active nematic fluid, the active-nematic MPCD approach coarse-grains the description of the bacteria suspension into a continuous-field hydrodynamic limit~\cite{saintillan2013}. 
The MPCD algorithm consists of two steps~\cite{MalevanetsKapral1999JCP-MPCD}. 

In step one, particles stream ballistically as 
\begin{align}
    \label{eq:streaming}
    \mpcdpos(t+\delta t) &=\mpcdpos(t)+\mpcdvel(t)\delta t  
\end{align}
for the streaming duration $\delta t$. 

In step two, a local multi-particle collision stochastically exchanges properties between the $N_c$ particles in each lattice-based cell $c$ of size $a$. 
Particles interact only within their local cells via collision operators. 
By constraining collision operators to conserve the appropriate quantities, MPCD reproduces hydrodynamics over sufficiently length and time scales~\cite{GompperIhle2009Book-MPCD}. 
In AN-MPCD, the collision operator exchanges momenta and orientations. 
The momentum collision operator $\collop_{i,c}$ is applied to each particle $i$ within cell $c$, generating a new velocity 
\begin{align}
    \mpcdvel(t+\delta t) &= \av{\vec{v}}_c(t) + \collop_{i,c},
    \label{eq:velColl}
\end{align} 
where $\av{\cdot}_c = \left(\sum_j^{N_c} \cdot_j \right)/N_c$ is the instantaneous cell average. 
A stochastic grid-shift applied to the lattice preserves Galilean invariance~\cite{GompperIhle2009Book-MPCD}.
The collision operator includes five contributions~\cite{kozhukhov2022}
\begin{align}
    \collop_{i,c} &= \collop^\text{and}_{i,c} + \collop^\text{ang}_{i,c} + \collop^\text{press}_{i,c} + \collop^\text{nem}_{i,c} + \collop^\text{act}_{i,c} ,
    \label{eq:totalcollop}
\end{align}
which represent the momentum-conserving Andersen-thermostat $\collop^\text{and}$, the angular-momentum conservation $\collop^\text{ang}$, the external driving pressure $\collop^\text{press}$, the nematic $\collop^\text{nem}$, and the active dipole $\collop^\text{act}$ terms. The first three are isotropic terms and the last two  are nematic and active-nematic terms. 

The Andersen-thermostatted operator~\cite{Gompper2007EPL} assigns each particle $i$ a  stochastic velocity $\mpcdvel^\text{ran}$ generated from a Maxwell-Boltzmann distribution of thermal energy $\kbt$ and ensures local momentum conservation through $\collop^\text{and}_{i,c} = \mpcdvel^\text{ran}-\av{ \vel^\text{ran} }_c$. 
Conservation of angular momentum $\angmom$ is achieved by applying a corrective angular velocity through $\collop^\text{ang}_{i,c}=(\tens{I}_c^{-1}\cdot\delta\angmom_\text{vel})\times\mpcdpos^{\prime}$. 
Here, $\tens{I}_c=m\sum^{N_c}_j\left(r_j^{\prime2}\identity-\vec{r}^\prime_j\vec{r}^\prime_j\right)$ is the local inertia tensor about the centre of mass, $\mpcdpos^\prime=\mpcdpos-\av{\vec{r}}_c$ are relative position vectors, and $\angmom_\text{vel}$ is the angular momentum due to the translational velocity of the point particles. 
An external driving force is included $\collop^\text{press} = g \hat{x} \  \delta t$, where $g$ is the external acceleration \fig{fig:passFlow}a)i). 
This external driving generates the pressure gradient $\grad P = -G\hat{x} = \dens g \hat{x}$ for density $\dens=m\av{ N_c }/a^d$ where $\av{\cdot}$ is the system average. 

Just as the momenta are stochastically exchanged in an isotropic fluid, the orientations must be stochastically exchanged for nematic MPCD (N-MPCD)~\cite{Shendruk2015SoftMatter-NMPCD}. 
Each particle has an orientation $\vec{n}_i(t)$ from which the local tensor order parameter and associated director $\dir_c$ and scalar nematic order parameter $S_c$ can be calculated within each cell $c$.
The director and scalar order are found from eigendecomposition of the nematic order tensor $\tensQ_c$ of each cell. 
During the collision operation, orientations are stochastically rotated by an orientation collision operator 
\begin{align}
    \dir_i(t+\delta t) &= \vec{\eta}_{i,c}\left(\dir_i(t);\dir_c(t)\right), 
    \label{eq:oriColl}
\end{align}
which samples from a Maier-Saupe distribution $\sim \exp{ \left(\mathcal{U} S_c (\dir_i \cdot \dir_c)^2 / \kbt \right) }$, centered about $\dir_c$ with a mean field interaction constant $U$~\cite{Shendruk2015SoftMatter-NMPCD}. 
The distribution makes use of a mean-field potential $\mathcal{U}$, representing the aligning potential between particles, which controls the isotropic-nematic transition and nematic elasticity~\cite{Shendruk2015SoftMatter-NMPCD}. 
The orientation and velocity fields are coupled via a two-step process~\cite{Shendruk2015SoftMatter-NMPCD}:

Firstly, shear alignment couples the orientations to velocity gradients via the Jeffery's equation for a slender rod with tumbling parameter $\lambda$ and hydrodynamic susceptibility $\chi$~\cite{Shendruk2015SoftMatter-NMPCD}. 
Each particle has overdamped rotational dynamics with balanced torque equations
\begin{align}
    \label{eq:torquebalance}
    \torque^{\mathrm{nem}}_i + \torque^{\mathrm{jeff}}_i + \torque^{\mathrm{diss}}_i &= 0 . 
\end{align}
The first torque $\torque^{\mathrm{nem}}_i = \gamma_R \dir_i \times \left(\frac{\delta \dir_i^\mathrm{col}}{\delta t}\right)$ is due to the orientational collision (\eqSI{eq:oriColl}), where $\gamma_R$ is the rotational friction coefficient. 
The second is due to hydrodynamic flows, $\torque^{\mathrm{HI}}_i = \gamma_R \dir_i \times \left(\frac{\delta \dir_i^{\mathrm{HI}}}{\delta t}\right)$, where the change in orientation is due to Jeffery's equation, $\frac{\delta \dir^{\mathrm{HI}}}{\delta t} = \chi \left[\vel_i \tens{\Omega} + \lambda \left(\dir_i \cdot \tens{E} - \dir_i\dir_i\dir_i : \tens{E} \right) \right]$, where $\tens{E}$ and $\tens{\Omega}$ are the strain rate and vorticity tensors. 
From \eqSI{eq:torquebalance}, the overdamped dissipitive term is $\torque^{\mathrm{diss}}_i = - \left(\torque^{\mathrm{nem}}_i + \torque^{\mathrm{jeff}}_i\right)$. 

Secondly, backflow coupling is applied by balancing torque on each particle with an equal and opposite torque on the fluid $\collop^\text{nem}_{i,c}= \gamma (\tens{I}_c^{-1}\cdot\delta\angmom_\text{ori})\times\mpcdpos^\prime$, for a viscous rotation coefficient $\gamma$
and change in angular momentum of  $\angmom_\text{ori} = \sum_{i}^{N_c}\torque^\mathrm{diss}_{i}\delta t$. 

The final term in the collision operator \eqSI{eq:totalcollop} is the active term. 
To induce active stresses within each cell, the active collision operator 
$\collop^\text{act}_{i,c}= \dir_c \left( \kappa_i - \av{ \kappa }_c\right) \left(\act \delta t / m\right)$ is applied~\cite{kozhukhov2022}.
This kicks half the particles forward and half backward ($\kappa_i\left(\mpcdpos^\prime; \dir_c\right) = \pm 1$) relative to $\dir_c$, generating a force dipole of strength $\act$~\cite{kozhukhov2023-ANSig}. 
Local, active forces arise from deformations of the orientation field as $\vec{f}_\text{act}\sim\act\grad\cdot\tensQ_c$ (\fig{fig:passFlow}a)ii).

\section{Simulation parameters}
\label{sctn:params}
All values are reported in MPCD units of point-particle mass $m$, collision-cell length $a$ and thermal energy $\kbt$. 
Relevant derived units in this study include time $\tau = a\sqrt{m/\kbt}$, activity with units of force $\kbt/a$, permeability with units of area $a^2$, pressure $\kbt / a^{d}$ in $d$ dimensions and pressure gradient $\kbt / a^{d+1}$. 
Numerically, each of the basic units is set to unity. 
As far as possible in the main text, quantities are non-dimensionalised by physically relevant scales, such as channel height $h=30$ or Poiseuille flow speed $\poiseuilleV=0.84$ for $\G=0.011$. 
The constant simulation parameters are $\delta t=0.1$, $\dens=20$, $\mathcal{U}=20$, $\lambda=2$, $\chi=1/2$, and $\gamma=0.01$. 
With these parameters, MPCD simulates linearised fluctuating nematodynamics with isotropic viscosity and elasticity~\cite{Hijar2019}. 
These parameters are chosen to match previous studies of AN-MPCD so that the bulk properties of the active nematic fluid match the established literature~\cite{kozhukhov2022}.
The director field is initialised normal to the channel walls. 
The activity is extensile, with values in the range $\act\in[0,0.09]$. 
External acceleration drives pressure gradients in the range $\G\in[0,0.022]$. 
Unless otherwise stated, the pressure-gradient is $\G=0.011$ and the activity is $\act=0.08$, which corresponds to an active length scale of $\actleng = 9.6 \simeq 10$. 
The active length scale $\actleng$ is measured as the mean separation between all topological defects in bulk turbulence as measured in a 2D system of size $100\times100$ with periodic boundary conditions.
The active length scale decays as $\actleng \sim \act^{-1/2}$ (\figSI{fig:actLength}), as expected from dimensional analysis of active nematics with a nematic Frank elasticity~\cite{Thampi2013}. 

\begin{figure}[h!]
    \centering
    \includegraphics[width=0.45\linewidth]{DefectSep.pdf}
    \caption{\textbf{Active length scale $\actleng$ as a function of activity $\act$.} 
    The bulk active length scale as measured in a  $100\times100$ system with periodic boundary conditions. 
    The length scale is measured as the mean separation distance between defects (of either charge) $\defleng$, which is the inverse square root of the number density of defects.
    The dotted line is a fit of $\defleng = a \av{\dens_n}^{-1/2}+b$ for $\av{\dens_n}$ the average number density of $\pm1/2$ defects, which is consistent with the expected power law decay of $\defleng \sim \actleng \propto \act^{-1/2}$.}
    \label{fig:actLength}
\end{figure}

All simulations undergo a warmup period of $0.5\times10^5$ steps, before a data collection period of $10^5$ steps. 
Data for each parameter set is averaged over 40 repeats (except \figSI{fig:finitesize} which uses 80 independent repeats). 
The instantaneous, fluctuating velocity field is the cell-averaged values $\vel\left(\vec{r},t\right)=\av{\vel}_c\left(\vec{r}_c,t\right)$. 
The longitudinal and transverse components of the velocity vector are $\vel = \vx\hat{x} + \vy\hat{y}$, respectively. 
Flow profiles and fluxes are averaged temporally $t$, longitudinally $x$ and/or transversely $y$, which we denote as $\av{\cdot}_{x,y,t}$, where subscripts $x,y,t$ are averaged over. 

\section{System setup}
\label{sctn:setup}
The active fluid and obstacles are confined within a channel of height $h=30$ with impermeable, no-slip walls with free nematic anchoring, unless otherwise stated. 
Simulations are performed in 2D. 
The channel length is $L=150$ with periodic boundary conditions. 
The porous medium is formed from impermeable, immobile, circular obstacles of radius $R=2$ with no-slip boundary conditions and free anchoring. 
Boundary conditions are achieved via collision-events rules between MPCD particles and solid surfaces: particle orientation $\dir_i(t)$ is unchanged but velocity $\mpcdvel(t)$ obeys bounce-back rules with the virtual particle method ensuring no-slip~\cite{Bolintineanu2012}. 
Free anchoring conditions are chosen here to mimic non-anchored motile bacteria and previous modeling of channel systems~\cite{shendruk2017dancing}. 
Comparisons of flow profiles for free and strong anchoring conditions are shown in \figSI{fig:anchCompare}. 
Strong anchoring is achieved by setting $\dir_i(t)$ to the surface normal (homeotropic) or tangent (planar) during collision events. 
Free and strong planar anchoring are qualitatively similar since active anchoring favours planar alignment at walls~\cite{blow2017motility}. 
 
The obstacles are randomly distributed according to a homogeneous probability distribution function (\fig{fig:passFlow}a)iii) but are not permitted to overlap with each other or the channel walls. 
This creates a sterically inaccessible layer of thickness $y/h \leq R/h=0.0\bar{6}$ that is devoid of obstacle centres (\fig{fig:passFlow}a)iii)~\cite{liu2007darcy}. 
An independent random configurations is generated for each of the 40 repeats. In each repeat, obstacles of radius $R$ are randomly placed under the constraint that they do not overlap with each other or with the channel walls.
Thus, the obstacle distribution is uniform between $R<y<h-R$ and zero elsewhere. 
The number of obstacles $N_\text{obs}$ sets the porosity $\por$, which is equivalent to $\por=1-\packfrac$, where $\packfrac=N_\text{obs} \pi R^2 / A$ is the obstacle area fraction for a total channel area of $A=hL$.
In this study, porosity is varied over $\por\in[0.67,1.0]$, where $\por=1.0$ is an empty channel. 
Using MC simulations, we find that there is a one sigma chance that the configuration of obstacles includes a set of obstacles that completely percolate across the channel height (with gaps $\leq 1$ between obstacles) for $\por=0.67$. 
Such a blockage necessitates zero flux since fluid can no longer elute, and so simulations are limited to $\por\geq0.67$. 

\section{Brinkman Equation}
\label{sctn:Brinkman}

These flows of passive nematics embedded in porous media and confined by boundary walls are described by the Brinkman equation~\cite{durlofsky1987analysis}
\begin{align}
    \left[ \frac{\partial^2}{\partial y^2} -\perm^{-1}\right]\av{\vx}_{x,t}({y}) &= -\frac{\G}{\visc},
    \label{eq:brinkman}
\end{align}
in which an effective friction coefficient term $-\vx/\perm$ accounts for Darcy drag and contributes to dissipation along with viscosity $\visc$. 
The interstitial permeability $\perm\left(\por\right)$ effectively smears hydrodynamic and nematic resistance due to the no-slip boundary conditions of the obstacles everywhere in space and is assumed to be isotropic and homogeneous. 

The interstitial permeability $\perm$ relates to the available paths through which fluid can advance. 
It is related to the porosity through $\perm \sim \por d_e^2$, where the immediate factor of $\por$ relates the total flow rate (superficial velocity) to the interstitial velocity (pore scale flows) and $d_e\left(\por\right) \sim A_\text{void}/P_\text{surf}$ is an effective hydraulic obstacle diameter for a void space $A_\text{void} \sim \por A$ and a total obstacle surface $P_\text{surf} \sim (1-\por) A$~\cite{berryman1987kozeny,probstein2005}. 
This results in the Kozeny–Carman equation $\perm = c \por^3(1-\por)^{-2}$, in which the coefficient $c$ accounts for tortuosity, surface roughness, particle shape, and other complications~\cite{graczyk2020predicting}.
Permeability is defined as $\perm=\lb^2$, where $\lb$ is the Brinkman pore size. This allows the Brinkman equation (\eqSI{eq:brinkman}) to be non-dimensionalised through $\lb$ and the Darcy velocity 
\begin{align}
    \darcyV &= \frac{\G\lb^2}{\visc} 
    \label{eq:darcy}
\end{align}
to be $\left[ \partial_{\tilde{y}\tilde{y}}-1\right]\av{\tilde{v}_x}_{x,t} = - 1$, where $\tilde{y}=y/\lb$ and $\tilde{v}_x=\vx/\darcyV$. 
Solving for no-slip boundary conditions gives
\begin{align}
    \driftND(\tilde{y}) &= 1-\cosh{\left(\tilde{y}\right)}+\tanh{\left( \frac{\tilde{h}}{2} \right)}\sinh{\left(\tilde{y}\right)} ,
    \label{eq:brinkmanfit}
\end{align}
where $\tilde{h} = h/\lb$. 
In the limit of $\tilde{h} \ll 1$, the average pore size is small compared to the channel height and so the flow profile approaches a constant value of $\darcyV$, corresponding to Darcy's plug-like flow. 
On the other hand in the limit $\tilde{h} \gg 1$, there are few obstacles and \eqSI{eq:brinkmanfit} approaches Poiseuille flow. 
The global average drift, defined as the mean flow over the entire channel
\begin{align}
    \avgdriftND &= 1-\left( \frac{2}{\tilde{h}} \right)\tanh{\left( \frac{\tilde{h}}{2} \right)} 
    \equiv \W\left(\tilde{h}\right),
    \label{eq:avdrift}
\end{align}
is nothing but Darcy's law with $\W\left(\tilde{h}\right)$ describing the correction due to a finite channel confinement. 
The hyperbolic tangent term smoothly switches the averaged drift from that expect for Darcy plug flow when $h \gg \lb$ to the mean of Poiseuille flow when  $\lb \gg h$. 
Similarly, the kinetic energy density is predicted to obey 
\begin{align}
    \keND &= \frac{ 2+\cosh(\tilde{h})-3\sinh(\tilde{h})/\tilde{h} }{ 1+\cosh(\tilde{h}) } .
    \label{eq:ke}
\end{align}
In the main text, these equations (\eqsSI{eq:brinkmanfit}{eq:ke}) are non-dimensionalised by Poiseuille speed $\poiseuilleV = \G h^2/12\eta$, rather than $\darcyV$, to avoid potentially obscuring the role of porosity. 
However, the form of the passive predictions remains the same with a dimensionless factor of $\darcyV/\poiseuilleV = 12/\tilde{h}^2$ (ratio of Darcy's speed and Poiseuille speed).

\section{Alternative forms of Active Darcy's Law}
\label{sctn:alternates}
The main text gives active Darcy's Law in terms of the effective active Darcy velocity (\eq{eq:activeDarcy}). 
However, the same information can be expressed in alternative equivalent forms. 
Firstly, the total drift can be written
\begin{align}
    \avgdrift &= \actdarcyV \W\left(h/\lb\right), 
    \label{eq:totalDrift}
\end{align}
which can be expanded (through \eq{eq:activeDarcy} and \eqSI{eq:avdrift}) to be
\begin{align}
    \avgdrift &= \darcyV \left[ 1+\frac{\act}{2\act^*}-\left(\frac{\act}{2\act^*}\right)^2 \right] \left[ 1-\left( \frac{2\lb}{h} \right)\tanh{\left( \frac{h}{2\lb} \right)}  \right].
    \label{eq:totalDriftExpanded}
\end{align}
where $\darcyV = \G\lb^2 / \visc$. 
The presence of the $\tanh{\left( h/ 2\lb \right)}$ factor accounts for the finite size of the channel. 
Alternatively, the modified Darcy's speed could be combined with the flow profile across the channel (\eqSI{eq:brinkmanfit}). 
The results can furthermore be written in terms of competitions between length scales, which makes the non-dimensionalised active Darcy's law flow profile across a finite channel
\begin{align}
    \frac{\flowpro\left(y/\lb\right)}{\darcyV} &= \left[ 1 + \frac{1}{2}\left(\frac{\lb}{\actleng}\right)^2 - \frac{1}{4}\left(\frac{\lb}{\actleng}\right)^4 \right] \left[1-\cosh{\left(\frac{y}{\lb}\right)}+\tanh{\left( \frac{h}{2\lb} \right)}\sinh{\left(\frac{y}{\lb}\right)} \right]. 
    \label{eq:fullFlow}
\end{align}
This form highlights how active Darcy's law arises as a competition between length scales. 
Firstly, Darcy's speed $\darcyV$ arises from the competition between the pressure gradient and the permeability $\perm=\lb^2$. 
Secondly, the active length scale $\actleng$ competes with the Brinkman length $\lb$ to produce the active enhancement. 
Finally, the finite size effects of the channel enter as a ratio of the channel height $h$ to the pore size $\lb$.

\section{Snapshots of density}
\label{app:densitySnapshot}

In systems of active particles, motility can induce phase separation into regions of high and low densities~\cite{cates2015} and giant-number fluctuations are a generically expected~\cite{Ramaswamy2003}. 
In bacterial systems, microbial motility leads to large number fluctuations~\cite{Zhang2010,liu2021}. 
However, continuum models of active nematics typically assume incompressibility~\cite{head2023}. 
In contrast, the AN-MPCD approach does not make this assumption. 
Rather, as a particle-based simulation method, AN-MPCD is not incompressible and naturally exhibits density fluctuations. 
This is one aspect of this coarse-grained method that makes it distinct from other numerical approaches. 
The density distributions in bulk turbulence tend to exhibit a widening the tails of the distribution with increased activity~\cite{kozhukhov2022}.  
The presence of the channel walls and obstacles limit variations in the density compared to bulk. 
The instantaneous density field $\dens$, as measured by the number of MPCD particles in each cell $\rho_c$, varies in space and time about the average value of $\av{\rho_c}=20$. 

\begin{figure}[h!]
    \centering
    \includegraphics[width=0.95\linewidth]{frame0059.pdf}
    \caption{\textbf{Snapshot of density fields in an AN-MPCD porous media simulation.} 
    The simulation corresponds to a purely active case (activity $\act=0.08$, pressure gradient $G=0$) through porous media of porosity $\por=0.916$. 
    As in all results presented here, the average number of particles per MPCD cell is $\av{ \rho_c }=20$, corresponding to a mass density of $\dens=m\av{ \rho_c }/a^d=20$. 
    }
    \label{fig:density}
\end{figure}

\section{Finite Size Effects}
\label{sctn:finiteSize}

A series of simulations demonstrates how the finite size of the channel impacts the results. 
When every length scale in the simulation is scaled up, the flow profile (normalised by Darcy's speed of $\darcyV =  \G\lb^2/\eta$) should be unchanged, according to \eqSI{eq:fullFlow}. 
This is indeed what is found when we scale up every length in the system by a factor of $2^{1/2}$ (\figSI{fig:finitesize}a). 
On the other hand, by increasing the system size while holding all other parameters fixed all terms in \eqSI{eq:fullFlow}, only the hyperbolic tangent term in $\W$ varies. 
Since all other terms remain unaltered, only the size of the near-wall area $\lb/h$ changes. 
Thus, the flow profiles should look the same but with a near wall region that shrinks relative to the channel height, making the profiles more and more plug-like, as confirmed by AN-MPCD simulations (\figSI{fig:finitesize}b). 
\begin{figure*}[h!]
    \centering
        \centering
            \centering
            \includegraphics[width=0.95\linewidth]{varH_HybFlow-act.pdf}
        \caption{\textbf{Role of finite size effects.}
        \textbf{(a)} 
        Scaling up all length scales by a constant factor leaves the results unchanged. 
        The blue curve is data from \fig{fig:diffJ} ($\act=0.08$; $\G=0.011$ and $\por=0.916$). 
        The green curve has increased all the length scales in the system by a factor of $\sqrt{2}$. The activity is halved to increase the active length scale by a factor of $\sqrt{2}$, the porosity is $0.947$ to obey the Kozeny–Carman equation, and the obstacle radius and channel height are multiplied by $\sqrt{2}$. 
        \textbf{(b)} 
        Increasing the channel height $h$ results in more plug-like flow. 
        Simulations at a porosity of $\por=0.916$ (corresponding Brinkman length of $\lb=3.68$), activity of $\act=0.08$ and pressure gradient $\G=0.011$. 
        The channel height of $h=30$ is the standard used in the manuscript.}
    \label{fig:finitesize}
\end{figure*}

\section{Effect of Anchoring}
\label{app:anchoring}

Free anchoring is the most relevant boundary condition for suspensions of planktonic bacteria  because there is no thermodynamic potential that favours particular alignments. 
That is, we neglect adhesion proteins expressed on the surface of planktonic cells that might lead to non-reversible attachment and orientational anchoring.
The picture is complicated because active stresses favour planar alignment at walls~\cite{blow2017motility}. 
A series of three simulations consider three extreme anchoring conditions (\figSI{fig:anchCompare}). 
Consistent with activity-induced planar anchoring, the average flow profile across the channel $\flowpro\left(y\right)$ is very similar for strong planar (\figSI{fig:anchCompare}; solid blue line) and free anchoring (\figSI{fig:anchCompare}; red dashed line). 
Strong homeotropic anchoring displays a flow profile with a greater magnitude than the planar case (\figSI{fig:anchCompare}; cyan dotted line). 
\begin{figure}[h!]
    \centering
    \includegraphics[width=0.45\textwidth]{anchCompare.pdf}
    \caption{
        \textbf{Effect of obstacle anchoring at porosity $\por=0.93$.} Comparison of flow profiles for an active fluid driven by a pressure gradient ($\act=0.08, \G=0.011$) for free (solid blue line), strong planar (dashed red) and strong homeotropic (dotted cyan) anchoring conditions on all obstacles and channel walls. 
        Position and flow profiles non-dimensionalised by channel height $h$ and Poiseuille flow $\poiseuilleV$, respectively. 
    }
    \label{fig:anchCompare}
\end{figure}

\section{Snapshots of hybrid active/pressure-driven flows}
\label{app:hybridSnapshots}

Snapshots of the hybrid pressure-driven and active system corresponding to \fig{fig:diffJ} are shown in \figSI{fig:hybSnapshot}. 
In each of these the activity is $\act=0.08$, and pressure gradient is $\G=0.011$ (from left to right).
The lowest porosity is $\por=0.93$ or permeability $\perm=17.5$ (\figSI{fig:hybSnapshot}a). 
This corresponds to \movie{mov:MidPhi-highAct}. 
This moderate porosity and high activity corresponds to fluxes below the maximum flux in \fig{fig:diffJ}c). 
The flow is primarily unidirectional flow, with localised regions of circulation. 
As the porosity is decreased ($\por=0.87$), the increased number of obstacles breaks up the streams (\figSI{fig:hybSnapshot}b). 
This produces higher tortuosity. 
Indeed as the porosity is further decreased to $\por=0.79$ (\figSI{fig:hybSnapshot}c), the flow must snake its way around clusters of obstacles. 
Significant portions of the flow are not parallel to the pressure gradient. 
\begin{figure}[h!]
    \centering
    \includegraphics[width=0.95\textwidth]{hybSnapshot.pdf}
    \caption{
        \textbf{Snapshots of hybrid flow.} Velocity field $\vel$ of an active nematic driven by a pressure gradient (activity $\act=0.08$, pressure gradient $G=0.011$) through porous media of porosity $\por$ coloured by speed. 
        \textbf{a)} $\por=0.97$. \textbf{b)} $\por=0.93$. \textbf{c)} $\por=0.87$. 
    }
    \label{fig:hybSnapshot}
\end{figure}

\clearpage
\section{Bulk Porous Systems}
\label{app:pbcs}

The simulations presented in the main text and other sections of the SI are of an active fluid and porous obstacles embedded within a 2D channel with impermeable no-slip walls (\sctnSI{sctn:setup}). 
In these systems, the effect of the walls is accounted for by the factor $\W$ (\eqSI{eq:avdrift}) and the system size effects have been considered in \figSI{fig:finitesize}. 
However, bulk porous systems can be simulated by replacing the impermeable, no-slip walls with periodic boundary conditions and allowing the obstacles to be placed anywhere within the system as long as they do not overlap. 
Here, we simulate a $150\times150$ system with $\G=0.011$ and $\perm=\lb^2=5.75$ ($\por=0.874$). 
A snapshot for $\act=0.08$ is shown in \figSI{fig:pbcs}a and activity is varied in the range $\act\in[0,0.09]$ in \figSI{fig:pbcs}b. 
The active contribution to the flux $\J$ is confirmed to be non-monotonic, increasing below $\act^*$ and then rapidly drop after $\act^*$ (\figSI{fig:pbcs}b). 
This is in agreement with the form of the active Darcy's law presented in the main text (\eq{eq:activeDarcy}). 
Furthermore, the optimal activity agrees with \fig{fig:diffJ}c for the same pressure gradient and permeability but in a channel of  height $h=30$. 
Indeed, the fit for the channel system is seen to agree with the bulk porous system .
No further fitting is required. 
This is because the factor of $\W$ in \eq{eq:actContr} completely accounts for the effect of the channel walls.  

\begin{figure}[h!]
    \centering
    \includegraphics[width=0.95\textwidth]{PBC-SI.pdf}
    \caption{
        \textbf{Active Darcy's law in a bulk porous system.} 
        \textbf{a)} Snapshot of the velocity field in a $150\times150$ system with periodic boundary conditions on all edges (activity $\act=0.08$, pressure gradient $G=0.011$, porosity $\por=0.874$) coloured by speed. 
        \textbf{b)} Active auxiliary contribution to the flux $\J$ as a function of activity $\act$ for $\perm/h^2=1.9\times10^{-2}$ ($\por=0.93$) with $\G=0.011$. 
        The non-monotonic dependence of $\J$ on $\act$ has a maximum at $\act^*=0.05$.
    }
    \label{fig:pbcs}
\end{figure}

\clearpage 
\section{Movie Captions}
\label{app:movies}

\begin{enumerate}
    \item 
    Active turbulent flow regime in an empty channel ($\por=1.0$) for activity $\act=0.08$ and no pressure gradient ($G=0$), with no-slip boundary conditions and free anchoring. The channel height is $h=30$, with periodic boundary conditions separated by a channel length of $L=150$. Arrows show velocity vectors, coloured by speed (in all movies). 
    \label{mov:Empty-turbulence} 
    \item 
    Purely active flow in a channel with porosity $\por=0.874$ ($\perm = 5.75$), activity $\act=0.08$, and no pressure gradient ($G=0$). Circular obstacles of radius $R=2$ coloured black (in all movies). The Brinkman pore size $\ell_B$, is smaller than the activity length scale $\actleng$, producing unidirectional flows in spontaneously chosen directions in parts of the channel. 
    \label{mov:HighPhi-streaming} 
    \item 
    Purely active flow in a channel of porosity $\por=0.791$ ($\perm=2.10$), activity $\act=0.08$, and no pressure gradient ($G=0$). Where local pore size matches the active length scale $\actleng$, recirculating vortices become trapped persisting for long times. 
    \label{mov:LowPhi-trapped} 
    \item 
    Hybrid active/driven flows in a porous medium of porosity $\por=0.930$ ($\perm=17.5$), activity $0.03$, and pressure gradient $G=0.011$ (from left to right). Such moderate porosity and low activity is below the maximum flux in \fig{fig:diffJ}c) and induces unidirectional flow with localised circulations due to activity. The pressure gradient dominates the behaviour. 
    \label{mov:MidPhi-lowAct} 
    \item 
    Hybrid active/driven flows in a porous medium of porosity $\por=0.930$ ($\perm=17.5$), activity $0.05$, and pressure gradient $G=0.011$ (from left to right). This is the flow behaviour for the maximum flux in \fig{fig:diffJ}c. Though vortices pattern the system, the flow is strongly biased by the pressure gradient and largely unidirectional. 
    \label{mov:MidPhi-peakAct} 
    \item 
    Hybrid active/driven flows in a porous medium of porosity $\por=0.930$ ($\perm=17.5$), activity $0.08$, and pressure gradient $G=0.011$ (from left to right). This is the flow behaviour in the high activity limit above the maximum flux in \fig{fig:diffJ}c. The flow dynamics within the porous medium are largely turbulent, consisting of disorderly and uncorrelated vortices that are only perturbed to stream in the direction of the pressure gradient. 
    \label{mov:MidPhi-highAct} 
    \item 
    Hybrid active/driven flows in a bulk porous medium of porosity $\por=0.874$ ($\perm=5.75$), activity $0.05$, pressure gradient $G=0.011$ (from left to right) and system size $150\times150$ with periodic boundary conditions on all sides. This is the flow behaviour for $\act^*$ in \fig{fig:diffJ}c. 
    \label{mov:PBC-peakAct} 
    \item 
    Hybrid active/driven flows in a bulk porous medium of porosity $\por=0.874$ ($\perm=5.75$), activity $0.08$, and pressure gradient $G=0.011$ (from left to right) and system size $150\times150$ with periodic boundary conditions on all sides. This is the high-activity flow behaviour of $\act>\act^*$. 
    \label{mov:PBC-highAct}
\end{enumerate}
\bibliography{references}